\newcommand*{\ARXIV}{}%
\ifdefined\ARXIV
\else
\fi

\ifdefined\ARXIV
\documentclass{article}
\usepackage{arxiv}

\else
\documentclass[sigconf, screen, nonacm]{acmart}
\fi

\AtBeginDocument{%
  \providecommand\BibTeX{{%
    \normalfont B\kern-0.5em{\scshape i\kern-0.25em b}\kern-0.8em\TeX}}}

\ifdefined\ARXIV
\else
\setcopyright{acmcopyright}
\copyrightyear{2021}
\acmYear{2021}
\acmDOI{10.1145/1122445.1122456}

\acmConference[ACM'21]{ACM '21: ACM Symposium on ACM Symposiums}{June 03--05, 2021}{Online, Earth}
\acmBooktitle{ACM '21: ACM Symposium on ACM Symposiums, June 03--05, 2021, Online, Earth}
\acmPrice{15.00}
\acmISBN{978-1-4503-XXXX-X/18/06}
\fi

\title{What We Measure in Mixed Reality Experiments}

\ifdefined\ARXIV

\author{Anthony Steed\\
Department of Computer Science\\University College London\\
\texttt{A.Steed@ucl.ac.uk}
}
\else
\author{Anthony Steed}
\email{A.Steed@ucl.ac.uk}
\orcid{0000-0001-9034-3020}
\affiliation{%
  \institution{Department of Computer Science, University College London}
  \streetaddress{66-72 Gower St}
  \city{London}
  \country{UK}
  \postcode{WC1E 6EA}
}

\fi

\ifdefined\ARXIV
\else

\fi

\begin{document}

\ifdefined\ARXIV
\maketitle
\else

\fi

\begin{abstract}
  
There are many potential measures that one might use when evaluating mixed-reality experiences. In this position paper I will argue that there are various stances to take for evaluation, depending on the framing of the experience within a larger body of work. I will draw upon various types of work that my team has been involved with in order to illustrate these different stances. I will then sketch out some directions for developing more robust measures that can help the field move forward.

\end{abstract}





\maketitle

\section{Introduction}

Mixed-reality (MR) experiences are somewhat unique as human-computer interfaces because they immerse the user within computer displays that replace or augment their senses. One class of virtual reality (VR) experiences provided by head-mounted displays (HMDs), is now a consumer product. Other classes of mixed-reality such as optical see-through augmented reality (AR) are available, though still relatively niche. Finally, classes such as video-mixed AR and projection-based VR are still mostly confined to laboratories. Given these different situations of use of MR and their different levels of maturity and specialism, we might expect the types of evaluation of system to be very different. There are MR systems being built to be used by millions of people (mostly games at this time), where there are MR systems that are unique in that they are engineered for the purpose of an evaluation (usually a controlled experiment).

In this position paper I illustrate different types of measurement by characterising the objective of the study within which they were used. In particular, I would argue that we should be sensitive to the objectives of the MR system, as artefact itself, or as a tool for use. Most of the work I will present was done for VR systems, but the observations would be applicable to a broader class of MR system.

\section{Measurements in Use}

As noted in the call for this workshop, a very wide variety of metrics have been used in MR studies. These range from biometrics of user reaction to specific events measured over milliseconds through to ethnographic studies of user engagement with MR over months. The question of what metrics are appropriate depends on your stance as a researcher. Are you interested in developing the MR systems themselves, in exploiting MR within a scientific content, in using MR as a tool, or exploring MR as a medium for communication? These are not comprehensive of the stances one might take, but they illustrate different themes of what role measurements have.

\subsection{MR as Novel HCI}

One characteristic of early work on VR, i.e. circa 1990-2000, is demonstrating why the technology was interesting to study as a novel form of HCI. Aside from the general novelty of the situation, as represented in popular media, one could argue that the notion of \emph{presence}~\cite{lombard_at_1997}, that is, that people behaved in virtual reality in some ways as we expect that they would in similar situations, was crucial as an early grounding of the promise of VR technology.  The literature on presence is vast, spanning a variety of disciplines and spawning specialist conference series. It covers measures, determinants, comparisons between systems and so on. Various presence questionnaires have been proposed, of which three have been utilised collectively by hundreds of studies (Witmer-Singer~\cite{witmer_measuring_1998}, Slater-Usoh-Steed~\cite{slater_depth_1994} and ITC-Sopi~\cite{lessiter_cross-media_2001}). I will not here argue about the validity or utility of these questionnaires, but today I would argue that they whether or not a HCI is high or low presence supporting is a very narrow view that doesn't reveal much. Those questionnaires emerged at a period where good quality VR was a difficult task: it was valid to ask whether a display engendered a feeling of "being there", because there was bad VR around (e.g. systems with high latency or narrow field of view) or the study was a comparison to a desktop display. 

My personal view is that there are two interesting directions here. The first is under-pinning the role of VR, and more generally MR, as embodied experience~\cite{kilteni_sense_2012} due to the sensori-motor contingencies~\cite{slater_place_2009}. This spurred a range of studies of what types of types of configuration of system support this embodied experience. This is an interesting direction because we can aim to understand the extent of implications of embodiment, and these might yield between system comparisons. For example, while this specific task is a bit too narrow as a general metric, because it is interactive, our own test of cognitive performance depending on embodiment~\cite{steed_impact_2016} suggests that there might be a series of measures of the implications of being immersed, and then being embodied in the environments.

A second direction is the range of perceptual phenomena that might only occur in immersive systems. That is, given the sensori-motor contingencies, are there sensory phenomena that are unique to the system, and how do we support these. Some of the underlying questions are technological drivers of MR technology such as resolution or field of view, but there are open questions such as whether any specific resolution or field of view is sufficient for the task. While you could embed a Snellen Chart in any MR system and use this to compare systems, there is a large space of other metrics that we could collectively use to compare different systems. 

\subsection{MR as Scientific Instrument}

Strongly related to perceptual metrics is the role of MR systems as instruments in other domains. This can cover domains ranging from perceptual neuroscience or sociology. They key features is that an MR system provides a controllable system that can reliably reproduce scenes between participants in experiments and that expectations about the real world can be probed by violating those expectations (e.g.~\cite{somervail_movement_2019} where expectations about gravity are manipulated in VR). Here the relationship between the domain of the study and the study of MR can be quite tight and yield useful insights in both directions. For example, studies in cognition where MR is used to create impossible spaces, can yield interesting insights into how to support navigation around virtual spaces. The community adopts measures from that community, and can use them as measures.

There are three interesting directions here. The first is that the capabilities of MR and thus the reasons why scientists tend to use them as research platforms, can give us unique insights into what metrics might be unique to MR as a HCI.  For example, an early paper from our lab looked at a depth version of the well known Posner effect (that the brain's visual attention system can be biased to attend and react more quickly to one hemisphere of vision), and this gave us hints about how to elaborate why stereo vision was important~\cite{maringelli_shifting_2001}. 

The second direction is that these collaborations provide new methods and techniques to evaluate systems. Insights from neuroscience, vision science, social science, etc. can inform us of what to expect in VR. For example, biometric signals are now commonly used in MR experiments as measures of responses.

The third direction is that knowledge of real behaviour and responses can give us goals for the development of MR technology because there is a phenomena that we wish to explore. 

\subsection{MR as Tool}

As interfaces, MR systems provide tools for us to interact. Over the past five years, enormous amounts of new design has been done in consumer applications. While most of these applications are games or entertainment, these still include tools for interaction, such as movement, manipulation and system interaction. Thus they are well within the scope of methods for analysis through various methods. 

Our main observation here is that the field of 3D user interaction has very broadly looked at the component technologies over the years and there are now hundreds of proposed interaction techniques based on using the wide variety of input controllers (e.g. see \cite{laviola_jr_3d_2017}).

A first direction here would be to look at revisiting the concept of testbed evaluation methods for MR~\cite{bowman_testbed_2001}. While there is value in defining characteristic tasks where a carefully-designed technique is superior (e.g. see \cite{steed_evaluating_2005}), with devices now targeted at consumer devices, it is desirable to see some conventions emerge that allow users to pick up and use applications quickly and be able to develop transferable skills between applications. The most obvious example here would be metrics to establish the most effective short-range travel technique as this a task that is common in most applications, but for which users have strong preferences.

A second direction is to develop better methods and metrics to inspect the usability of MR applications. It is discussed in early work in the field, that usability methods designed for 2D applications are difficult to extend to immersive systems, because of the freedom of movement, and the broad range of tasks that immersive systems support. However, there are specific concerns, ranging from avoiding simulator sickness through to measuring the interface's exploitation of the surrounding nature of the task that might be amenable to direct evaluation.

A third direction is to start to share data in a more systematised way. Each user has a different experience with the usability of the system. Previously several proposals have been made to create formats to log and share data from MR systems, for the purpose of analyse of user behaviour and replaying of events to explore specific events~\cite{murgia_tool_2008, steptoe_multimodal_2012}.  It is a good time to revisit these concepts because the tools, such as game engines and device support, are now much more mature. 

\subsection{MR as Communication}

The previous sections have considered MR as mostly a single-user experience and thus have focused on perception and interaction. My own opinion is within the potential applications of MR, communication, both synchronous and asynchronous, is the most promising opportunity. Synchronous communication has long been one of the motivating applications for MR. Based on the observation that something is missing from 2D representations, a huge range of work has looked at avatar-based communication or real-time reconstruction of people. While simple questionnaires can get at some aspects of feeling as if one is with others, our own work has focused on specific measurable traits such as leadership~\cite{steed_leadership_1999} or trust~\cite{pan_impact_2017}. Thus the first direction for future work is to establish some common metrics that can be used to establish the quality of communication and interaction between people. This could span from conversational fluidity through to establishment of empathetic stances. This might go hand-in-hand with standardised social tasks or even environments.

The second direction is how users adapt to the social environments over time, and what features they expect. Longitudinal studies have a role for studying MR as tools, but we mention it here because it is the social MR tools that are the most flexible, in that they commonly support user editing for the purpose of sharing, especially in their avatar representations (e.g. \cite{moustafa_longitudinal_2018}). We expect that new tools and metrics will be developed to study not only how effective communication is at any point in time, but how it evolves over time.

A third direction is less well defined, but notes that content creation, sharing, adaptation and re-use is well studied in other media. Thus there is a role for studying asynchronous MR communication and its effectiveness. This might be in online worlds, or situated in the real-world using the spatial anchors now provided to several AR systems. This blends into the effectiveness of MR as a medium for content creation, and the effectiveness of communication using creative artefacts, not just real-time communication. While there is an emerging critique of MR pieces out there, much of it that will hopefully remain beyond metrics, there are skills and crafts to be understood as MR is more broadly used. 
 
\section{Conclusions}

In our MR studies we have used dozens of different measurements of experiences, from questionnaires through biometrics to ethnographic methods. Each method has been chosen because of the particular question behind the work: are we studying the effectiveness of a user interface, the user reaction to a new rendering method or a subtle choice in the design of an interactive story-telling piece.  In this position paper I've outlined some of the directions we have been looking at when discussing how to develop new measures ni my group.

To finish, we would identify that the most effective methods and metrics are those that are easily used by other experimenters. Sometimes this is effected just by publishing, as with questionnaires. But there is a growing need for more standards and conventions to allow us to share methods and results so that we can be more open and transparent in recording our experiments, but also facilitating reproduction or re-analysis.


\ifdefined\ARXIV
\section*{Acknowledgements}
\else
\begin{acks}
\fi

To the many members of the Virtual Environments and Computer Graphics lab at UCL.

\ifdefined\ARXIV
\else
\end{acks}
\fi


\ifdefined\ARXIV
\bibliographystyle{unsrt}
\else
\bibliographystyle{ACM-Reference-Format}
\fi
\bibliography{sample-base}



\end{document}